%% file: commodities.tex
\documentclass[12pt]{article}
\usepackage{graphpap, graphicx, float, tabularx, array, setspace, natbib, multirow}
\usepackage[multiple]{footmisc}
\input{mathdef}

\long\def\symbolfootnote[#1]#2{\begingroup
\def\thefootnote{\fnsymbol{footnote}}\footnote[#1]{#2}\endgroup}
\numberwithin{equation}{section}
\theoremstyle{plain}

\renewcommand{\baselinestretch}{1.3}

\newcommand\blfootnote[1]{%
  \begingroup
  \renewcommand\thefootnote{}\footnote{#1}%
  \addtocounter{footnote}{-1}%
  \endgroup
}

\begin{document}

\centerline{\bf \LARGE{The Rank Effect for Commodities}}

%\vskip 12pt

%\centerline{\bf \LARGE{}}

\blfootnote{The views expressed in this paper are those of the authors alone and do not necessarily reflect the views of the Federal Reserve Bank of Dallas or the Federal Reserve System. All remaining errors are our own.}

\vskip 50pt

\centerline{\large{Ricardo T. Fernholz\symbolfootnote[2]{Claremont McKenna College, 500 E. Ninth St., Claremont, CA 91711, rfernholz@cmc.edu} \hskip 90pt Christoffer Koch\symbolfootnote[3]{Federal Reserve Bank of Dallas, 2200 North Pearl Street, Dallas, TX 75201, christoffer.koch@dal.frb.org}}}

\vskip 3pt

\centerline{\hskip 17pt Claremont McKenna College \hskip 57pt Federal Reserve Bank of Dallas}

\vskip 45pt

%\centerline{\large{First Draft: January 25, 2012}}

\vskip 5pt

\centerline{\large{\today}}

\vskip 50pt

\renewcommand{\baselinestretch}{1.1}
\begin{abstract}
We uncover a large and significant low-minus-high rank effect for commodities across two centuries. There is nothing anomalous about this anomaly, nor is it clear how it can be arbitraged away. Using nonparametric econometric methods, we demonstrate that such a rank effect is a necessary consequence of a stationary relative asset price distribution. We confirm this prediction using daily commodity futures prices and show that a portfolio consisting of lower-ranked, lower-priced commodities yields 23\% higher annual returns than a portfolio consisting of higher-ranked, higher-priced commodities. These excess returns have a Sharpe ratio nearly twice as high as the U.S. stock market yet are uncorrelated with market risk. In contrast to the extensive literature on asset pricing factors and anomalies, our results are structural and rely on minimal and realistic assumptions for the long-run behavior of relative asset prices.
%stable relative asset price distribution

%We show that a portfolio consisting of higher-ranked, more expensive commodities yields significantly lower returns than a portfolio consisting of lower-ranked, less expensive commodities over the 1980 -- 2015 period.  We confirm the existence of a similar size effect for commodities using historical data on global commodity prices from 1882 -- 1913, as predicted by our econometric theory.
\end{abstract}
\renewcommand{\baselinestretch}{1.3}

%This rank-based predictability is inconsistent with some versions of the efficient markets hypothesis, but it is also a necessary econometric consequence of a plausible relative asset price distribution, regardless of the risk and liquidity properties of those assets.

\vskip 50pt



\vfill

\pagebreak

\section{Introduction}

Many asset pricing anomalies have been documented in different asset markets. In response, as many as 300 different factors have been proposed as potential drivers of these anomalies \citep*{harvey/liu/zhu:2016}. A structural econometric explanation that relies on minimal, robust, and realistic assumptions for the long-run behavior of relative asset prices has not yet been proposed, however.

We show that rank-based, nonparametric econometric methods along with minimal and realistic economic assumptions predict the presence of a rank effect in many asset markets. As long as asset prices are normalized in an economically meaningful way, then they can be ranked. If the distribution of those normalized prices is stationary, then lower-ranked, lower-priced assets must necessarily have their prices grow more quickly than higher-ranked, higher-priced assets. In other words, a rank effect will exist.

From 2010 -- 2015, an equal-weighted portfolio of lower-ranked commodity futures earned an average excess return of 23.2\% per year over an equal-weighted portfolio of higher-ranked commodity futures. Over this same period, the Sharpe ratio of this excess return is nearly twice as high as for the U.S. stock market, yet the correlation between this daily excess return and the daily return on the Russell 3000 is 0.10. In 2015, for example, this excess return was 24.2\% even as the Russell 3000 generated a return of 4.1\%. In other words, we uncover a tradable rank effect for commodities that is large, significant, and uncorrelated with common measures of market risk. We confirm the presence of this rank effect using monthly spot commodity prices spanning more than a century.

%and the yield on 10-year U.S. treasuries was ??\%

The rank-based, nonparametric methods we apply in this paper are valid for a broad range of processes for individual commodity price dynamics. In particular, these processes may exhibit growth rates and volatilities that vary across time and individual commodity characteristics. Our econometric theory, which was first introduced to economics by \citet{Fernholz:2016c}, is well-established and the subject of active research in statistics and mathematical finance.\footnote{A growing and extensive literature, including \citet*{Banner/Fernholz/Karatzas:2005}, \citet{Pal/Pitman:2008}, \citet*{Ichiba/Papathanakos/Banner/Karatzas/Fernholz:2011}, and \citet{Shkolnikov:2011}, among others, analyzes these rank-based methods.} Although we do not commit to a specific model of economic behavior in this paper, the generality of our methods implies that our econometric framework is applicable to both rational \citep{sharpe:1964,Lucas:1978a,Cochrane:2005} and behavioral \citep{shiller:1981,debondt/thaler:1989} theories of asset price dynamics.

While the price dynamics of individual commodities are flexible, our econometric results are disciplined by the asymptotic properties of joint price dynamics. It is in this sense that our results are structural. We show that a stationary distribution of relative commodity prices requires a mean-reversion condition. Specifically, a stationary distribution exists only if the returns of higher-ranked, higher-priced commodities are on average lower than the returns of lower-ranked, lower-priced commodities. This prediction relies only on the properties of relative commodity price dynamics, and is supported by the data. Indeed, using multiple data sets of spot and futures commodity prices spanning sub-periods ranging from 1882 to 2015, we demonstrate a rank effect in which higher-ranked commodities systematically underperform lower-ranked commodities as predicted by our econometric theory.

How can commodity prices be appropriately normalized, compared, and ranked? We operationalize the notion of commodity rank by normalizing prices to a common starting value. In subsequent periods, commodities are ranked by comparing their normalized prices, so that in each period the \emph{low} (L) commodities are the lowest-ranked and lowest-priced and the \emph{high} (H) commodities are the highest-ranked and highest-priced. Our results show that simple portfolios that invest equal dollar amounts in the lower-ranked commodities consistently yield significantly higher returns than portfolios that invest equal dollar amounts in the higher-ranked commodities. Specifically, a \emph{low-minus-high} (LMH) investment strategy for commodities generates annual excess returns of 23.2\% despite almost no correlation with stock market returns.

This excess return is exactly what is predicted by our econometric framework. Furthermore, it is not at all obvious how such an excess return can be arbitraged away. After all, regardless of the rationality or irrationality of investors, commodity prices must always fluctuate and differ in a way that allows them to be ranked. This striking implication raises fundamental and difficult questions about the true meaning of market efficiency \citep{Fernholz:2016d}.

Over all sample periods we consider in this paper, the cross-sectional data are consistent with a stationary distribution of relative commodity prices. Beyond that, standard economic arguments support the presence of such stationarity in any sample. The fundamental economic notion of substitutability suggests that strong forces ensure the long-run stability of relative commodity prices and prevent any one commodity price from growing arbitrarily larger than all other commodity prices. One of the contributions of this paper is to link this economically reasonable and empirically realistic asymptotic property of the relative commodity price distribution to the surprising implication that there exist large rank effect excess returns.

In contrast to the large and growing literature on asset pricing anomalies, we present a flexible econometric theory that predicts the rank effect that we observe. This contribution of our paper is novel, and it implies that there is nothing anomalous about the anomaly that we uncover. The structural econometric foundation for our results is notably absent from both the anomalies and factor pricing literatures.

One novel critique of the extensive list of asset-pricing anomalies and factors that have been proposed focuses on invalid or incomplete inference about risk-premia parameters. \citet{bryzgalova:2016}, for example, shows that standard empirical methods applied to inappropriate risk factors in linear asset pricing models can generate spuriously high significance. In such cases, the impact of true risk factors could even be crowded out from the models. \citet{novy-marx:2014} provides a different critique, demonstrating that many supposedly different anomalies are potentially driven by one or two common risk factors. In other words, the extensive list of anomalies and factors proposed by the literature overstates their true number. Because our findings are founded in a well-established, nonparametric econometric theory that makes ex-ante testable predictions, they are not subject to such ex-post critiques. Indeed, we predict exactly the commodities rank effect anomaly that we find across different centuries and frequencies.

The rank effect for commodities that we uncover is conceptually similar to the well-known size effect for stocks. \citet{Banz:1981} was the first to show the tendency for U.S. stocks with high total market capitalization, big stocks, to underperform U.S. stocks with low total market capitalization, small stocks. This observation gave rise to the well-known size risk factor (SMB) first proposed by \citet{Fama/French:1993}.\footnote{For an extensive summary of the literature, see \citet{vandijk:2011}.} Our econometric theory states that the prices of higher-ranked assets must necessarily grow more slowly than the prices of lower-ranked assets, as long as the relative price distribution is stationary. If this methodology is applied to the size of stocks, then our theory predicts that higher-ranked, bigger stocks will yield lower capital gains than lower-ranked, smaller stocks on average over time. In other words, there will be a size effect.

In addition to our results for commodity futures prices, we also confirm the existence of a rank effect for spot commodity prices from 1980 -- 2015 and from 1882 -- 1913.\footnote{This latter rank effect is uncovered using a new data set of historical monthly spot commodity prices from \citet{Fernholz/Mitchener/Weidenmier:2016}.} We show that from 1980 -- 2015, an equal-weighted portfolio of lower-ranked commodities earns an average excess return of 6.4\% per year over an equal-weighted portfolio of higher-ranked commodities.\footnote{Furthermore, \citet{Fernholz:2016d} shows that the correlation between these rank effect excess returns for commodities and Russell 3000 stock returns is -0.13 during this same period.} Similarly, from 1882 -- 1913, an equal-weighted portfolio of lower-ranked commodities earns an average excess return of 16.5\% per year over an equal-weighted portfolio of higher-ranked commodities. Although these excess returns are not tradable, it is nonetheless notable that such a large and significant rank effect for commodities exists across multiple data sets that cover multiple centuries. Of course, this finding is not surprising given the predictions of our econometric theory. After all, there are good reasons for the distribution of relative spot commodity prices to be stationary much like the distribution of relative commodity futures prices, a fact that is supported by the econometric analysis of \citet{Fernholz:2016c}.

The rest of the paper is organized as follows. Section \ref{sizeEffect} presents our main results on the rank effect using both daily commodity futures prices and monthly spot commodity prices across two centuries. Section \ref{econom} develops our rank-based, nonparametric econometric theory and discusses the prediction that higher-ranked, higher-priced assets will underperform lower-ranked, lower-priced assets on average over time.  Section \ref{conclusion} concludes.

\vskip 50pt

\section{The Rank Effect} \label{sizeEffect}

Using both spot and futures commodity prices, we uncover a large and economically significant rank effect in which lower-ranked, lower-priced commodities outperform higher-ranked, higher-priced commodities. For commodity futures, we show that these predictable and tradable excess returns are practically uncorrelated with market risk. We also show that similar rank effect excess returns exist for spot commodity prices spanning more than a century.

\subsection{The Rank Effect for Commodity Futures} \label{futures}
We use daily data on the future price of 30 commodities from January 5, 2010 to January 14, 2016.\footnote{These commodities are aluminum, soybean oil, cocoa, corn, cotton, ethanol, feeder cattle, gold, copper, heating oil, ICE UK natural gas, orange juice, coffee, lumber, live cattle, lead, lean hogs, NYMEX natural gas, nickel, oats, platinum, rough rice, sugar, soybean meal, silver, soybeans, wheat, WTI crude oil, gasoline, zinc.} These data, which were obtained from TickWrite, report the price at 4pm GMT on each trading day for each commodity one month in the future. Because commodities are sold in different units and hence their prices cannot be compared in an economically meaningful way, it is necessary to normalize these prices by equalizing them in the initial period. This normalization permits these commodity futures prices to be compared and ranked.

\begin{figure}[ht]
\begin{center}
\vspace{-30pt}
\hspace{-15pt}\scalebox{0.63}{ {\includegraphics{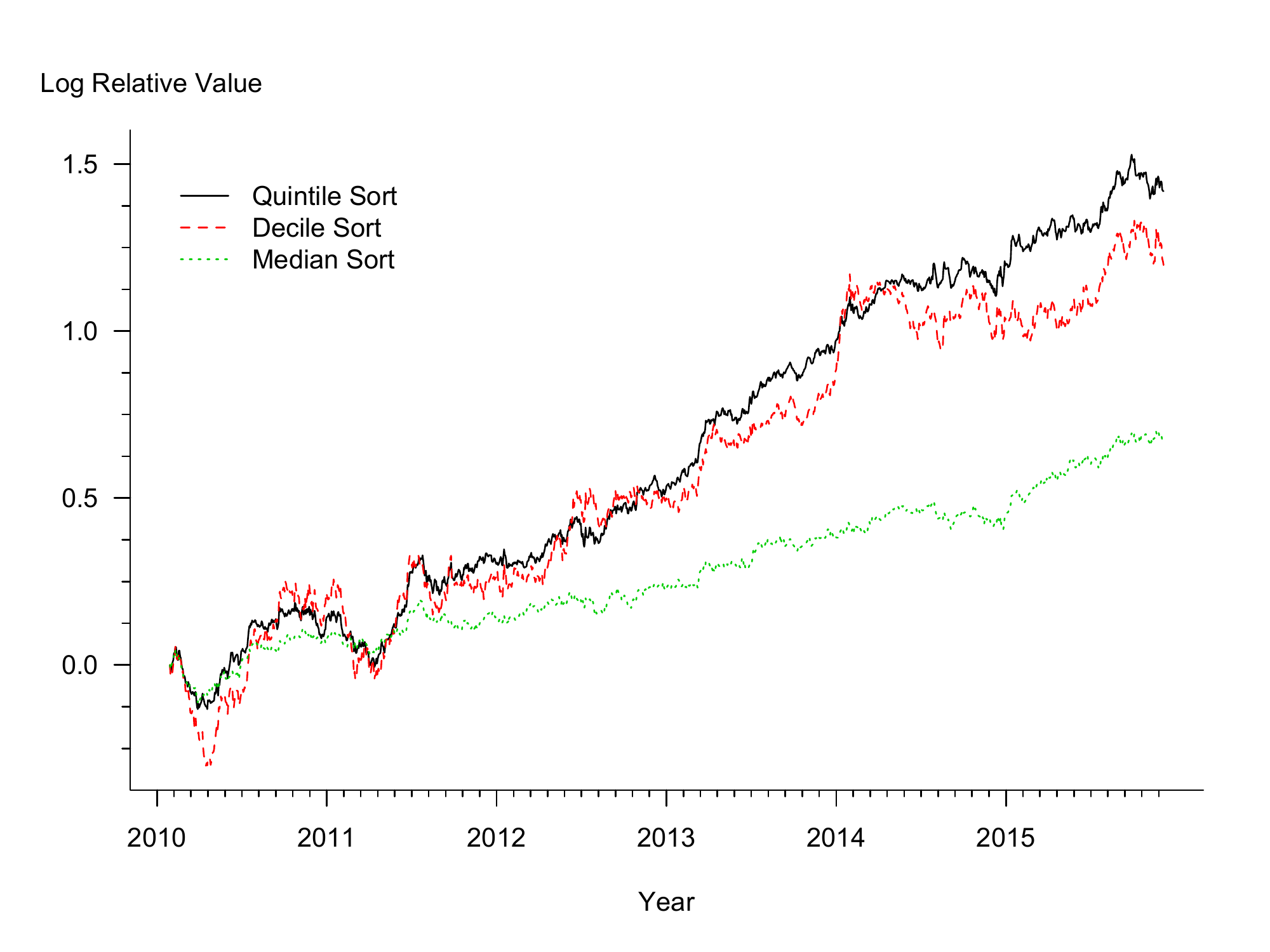}}}
\end{center}
\vspace{-24pt} \caption{Log return of lower-ranked (bottom decile, quintile, and half) commodity futures portfolios relative to higher-ranked (top decile, quintile, and half) commodity futures portfolios, 2010 -- 2015.}
\label{sizeFigFutures1}
\end{figure}

In Figure \ref{sizeFigFutures1}, we plot the log excess returns of three different portfolios of lower-ranked, lower-priced commodity futures over portfolios of higher-ranked, higher-priced commodity futures for 2010 -- 2015. These portfolios place equal weights on each commodity, so that the quintile sort reports the excess return of the six lowest-ranked (bottom 20\%) commodity futures over the six highest-ranked (top 20\%) commodity futures. All portfolios are rebalanced daily. In other words, the normalized futures prices are ranked on each day, and the quintile sort corresponds to the excess return of an investment of equal dollar amounts in each of the bottom six daily-ranked commodity futures over an investment of equal dollar amounts in each of the top six daily-ranked commodity futures. Commodities are ranked by comparing their normalized prices, so that each day the \emph{low} (L) commodities are the lowest-priced and the \emph{high} (H) commodities are the highest-priced. Finally, we wait 20 days after the start date of January 2010 to begin trading so that the rankings of commodities are meaningful and the distribution of relative commodity prices has time to approach a stationary distribution.\footnote{We discuss stationarity of relative commodity prices and its implications in Section \ref{econom}.} The median and decile sorts are constructed similarly to the quintile sort, but with cutoffs of fifteen ranks (50\%) and three ranks (10\%), respectively.

Figure \ref{sizeFigFutures1} clearly shows that that all three of these \emph{low-minus-high} (LMH) excess returns are large and economically significant. Indeed, for the quintile sort, the lower-priced, lower-ranked commodity futures portfolio generates an average annual excess return of 23.2\% over the higher-priced, higher-ranked commodity futures portfolio. Furthermore, the Sharpe ratio of these excess returns is nearly twice as high as the Sharpe ratio for the Russell 3000 stock index over the same 2010 -- 2015 period. The decile sort excess returns are similar to the quintile sort. The median sort average annual excess return is smaller, at 11.1\%, but it is also considerably less volatile, as Figure \ref{sizeFigFutures1} shows.

\begin{figure}[ht]
\begin{center}
\vspace{-5pt}
\hspace{-15pt}\scalebox{0.63}{ {\includegraphics{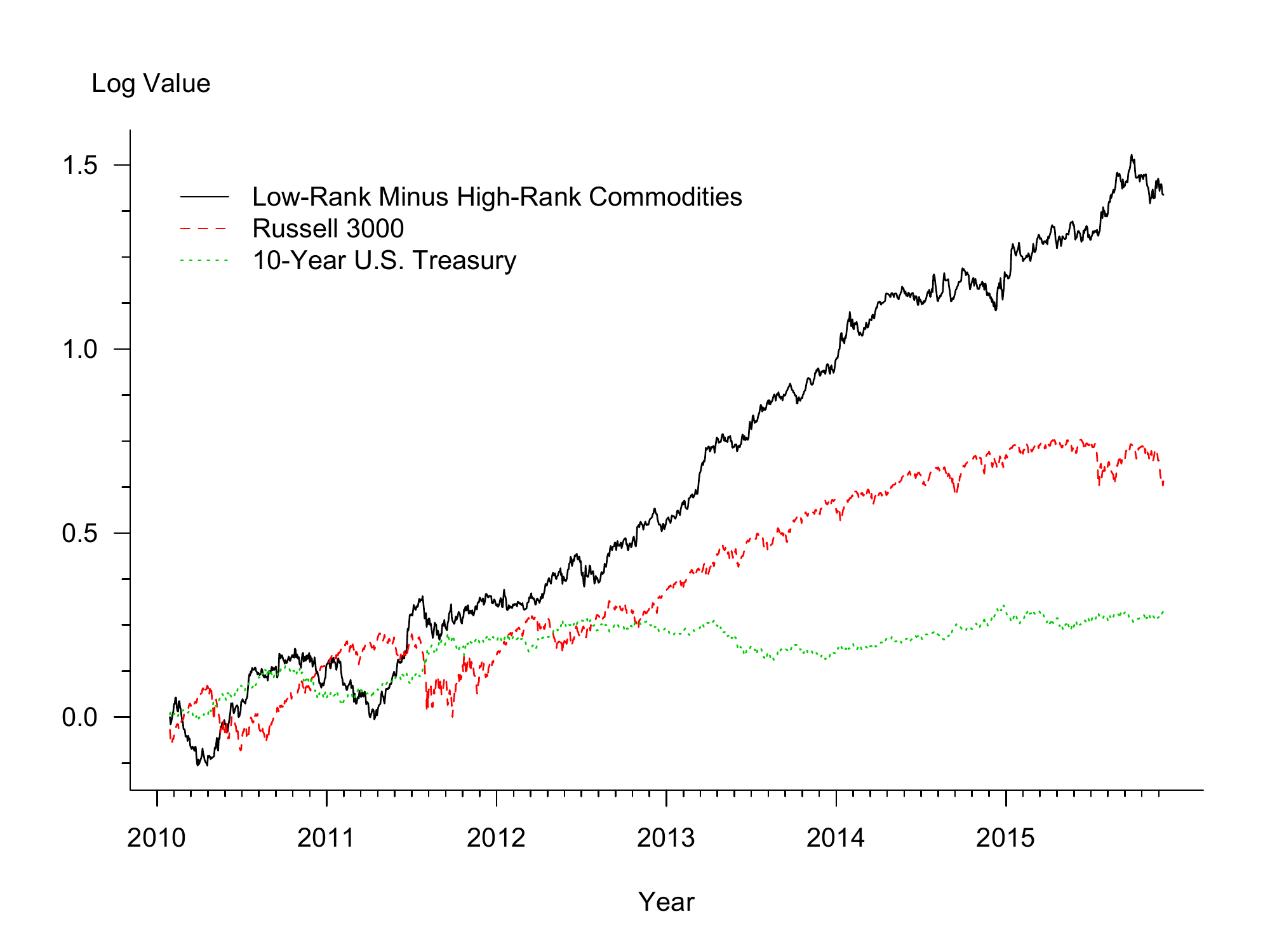}}}
\end{center}
\vspace{-24pt} \caption{Log relative return of lower- versus higher-ranked (quintile sort) commodity futures portfolios, log return of Russell 3000, and log return of 10-year U.S. treasury constant maturity total return index, 2010 -- 2015.}
\label{sizeFigFutures2}
\end{figure}

The excess returns from this rank effect for commodities are also practically uncorrelated with market risk. In Figure \ref{sizeFigFutures2}, we plot the log excess return of an equal-weighted portfolio of lower-ranked, lower-priced commodity futures over an equal-weighted portfolio of higher-ranked, higher-priced commodity futures for 2010 -- 2015 together with the log value of the Russell 3000 index and the log value of a 10-year U.S. treasury constant maturity total return index. The LMH excess returns shown in this figure correspond to the quintile sort from Figure \ref{sizeFigFutures1}. As Figure \ref{sizeFigFutures2} shows, the rank effect for commodities generates excess returns that substantially outperform both the U.S. stock market and 10-year U.S. treasuries over a period during which both of those investments performed fairly well. The LMH excess returns for commodities shown in the figure also do not appear to be correlated with market risk. Indeed, in 2015, the LMH excess return for commodities was 24.2\% even as the Russell 3000 generated a total return of only 4.1\%.

Table \ref{regressionsTab} reproduces the results in Figures \ref{sizeFigFutures1} and \ref{sizeFigFutures2}, and confirms that the rank effect for commodities is not explained by market risk. The table reports the regression results for daily excess returns of equal-weighted, lower-ranked commodity futures portfolios over equal-weighted, higher-ranked commodity futures portfolios for all fifteen possible rank cutoffs from 2010 -- 2015. In particular, the table shows the intercept and coefficient of a regression of daily LMH excess returns on daily Russell 3000 returns for LMH excess returns of the bottom 1-15 ranked commodity futures relative to the top 1-15 ranked commodity futures. For practically all rank cutoffs, Table \ref{regressionsTab} shows that these LMH abnormal excess returns are highly statistically significant and essentially unexplained by market risk. Indeed, for the quintile sort (rank cutoff of six), the intercept of 9.24 basis points is more than three times larger than the standard error and the coefficient for Russell 3000 daily returns is only 0.104. The median sort (rank cutoff of fifteen) yields similar results, only with a lower intercept and a lower coefficient for market returns. It is only for the very highest rank cutoffs that the intercepts lose their significance, but this is not surprising given that such portfolios hold only one or two commodity futures at a time.

Figure \ref{sizeFigFutures3} plots the log value of an equal-weighted portfolio that invests in the bottom fifteen commodity futures together with the log value of an equal-weighted portfolio that invests in the top fifteen commodity futures for 2010 -- 2015. According to the figure, the rank effect excess returns are split fairly evenly between growth in the prices of the lowest-ranked commodity futures and declines in the prices of the highest-ranked commodity futures. Figure \ref{sizeFigFutures4} plots the log value of an equal-weighted portfolio that invests in the bottom fifteen commodity futures relative to the log value of an equal-weighted portfolio that invests in the top fifteen commodity futures for 2010 -- 2015. These log excess returns correspond to the median sort reported in Figure \ref{sizeFigFutures1} and Table \ref{regressionsTab} (rank cutoff fifteen). We show in Section \ref{econom} that these abnormal returns for commodity futures prices are not surprising and are in fact a necessary consequence of the stationarity of the distribution of relative commodity futures prices.

%For quintile sort, average daily excess returns of 9.78 basis points and a correlation of 0.1 with the total return on the Russell 3000 index from 2010 -- 2015.

\subsection{The Rank Effect Across Two Centuries} \label{century}
In Section \ref{futures}, we uncovered a large and economically significant rank effect for daily commodities futures prices over the 2010 -- 2015 period. In this section, we show that the structural features of relative price stationarity and mean reversion that drive this generalized size effect are also present in monthly spot commodity prices sampled across two centuries. We use monthly data on the spot price of 22 commodities from 1980 -- 2015 obtained from the Federal Reserve Bank of St. Louis (FRED), and monthly data on the spot price of fifteen commodities from 1882 -- 1913 recorded by \citet*{Fernholz/Mitchener/Weidenmier:2016} using the ``Monthly Trade Supplement'' of the \emph{Economist}.\footnote{For 1980 -- 2015, these commodities are aluminum, bananas, barley, beef, Brent crude oil, cocoa, copper, corn, cotton, iron, lamb, lead, nickel, orange, poultry, rubber, soybeans, sugar, tin, wheat, wool (fine), and zinc. For 1882 -- 1913, these commodities are barley, beef, copper, cotton, hemp, jute, lead, mutton, oats, rice, silver, sugar, tea, tin, and wheat.}

\begin{figure}[ht]
\begin{center}
\vspace{-30pt}
\hspace{-15pt}\scalebox{0.66}{ {\includegraphics{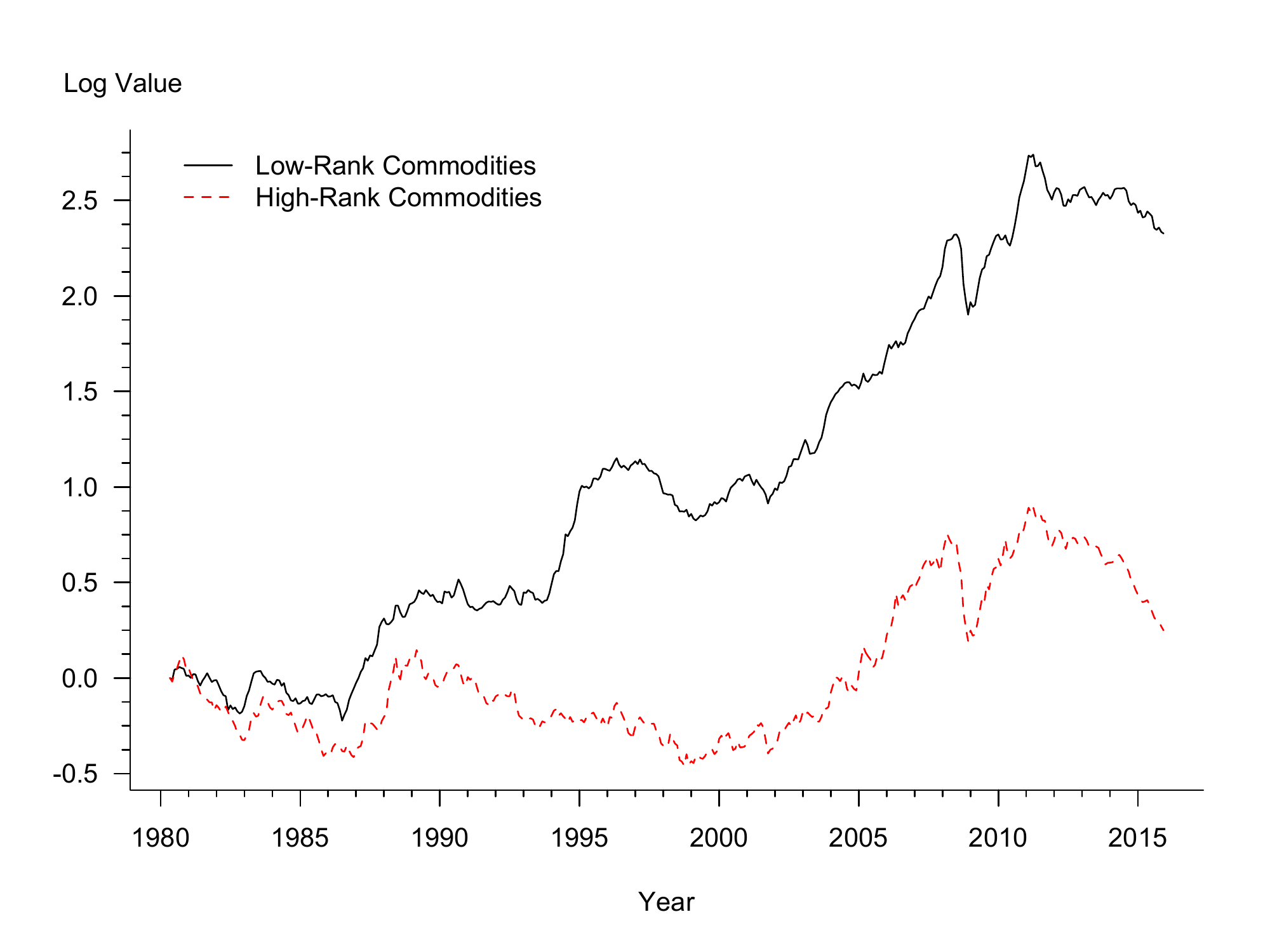}}}
\end{center}
\vspace{-24pt} \caption{Log returns for lower-ranked (bottom half) commodities and higher-ranked (top half) commodities portfolios, 1980 -- 2015.}
\label{sizeFigRecent1}
\end{figure}

\begin{figure}[ht]
\begin{center}
\vspace{-5pt}
\hspace{-15pt}\scalebox{0.66}{ {\includegraphics{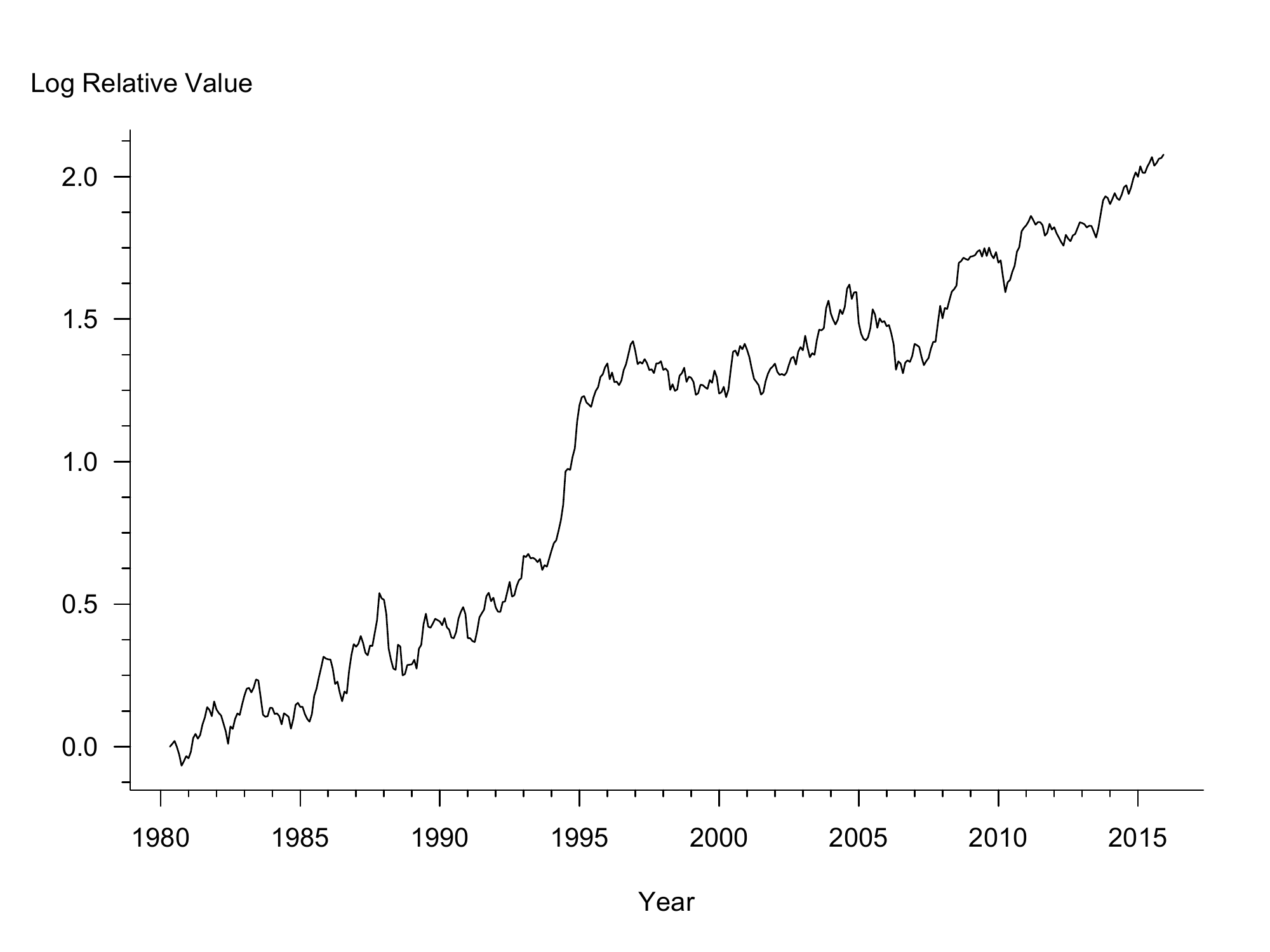}}}
\end{center}
\vspace{-24pt} \caption{Log return of lower-ranked (bottom half) commodities portfolio relative to higher-ranked (top half) commodities portfolio, 1980 -- 2015.}
\label{sizeFigRecent2}
\end{figure}

In Figure \ref{sizeFigRecent1}, we plot the log value of an equal-weighted portfolio that invests in the bottom eleven commodities together with the log value of an equal-weighted portfolio that invests in the top eleven commodities for 1980 -- 2015. Figure \ref{sizeFigRecent2} plots the log value of the bottom-eleven commodities portfolio relative to the top-eleven commodities portfolio from Figure \ref{sizeFigRecent1}. Similar to the commodity futures portfolios constructed in Section \ref{futures}, we wait five months after the start date of January 1980 to begin trading so that the distribution of relative commodity prices has time to approach a stationary distribution and the rankings of commodities are meaningful. 

\begin{figure}[ht]
\begin{center}
\vspace{-30pt}
\hspace{-15pt}\scalebox{0.66}{ {\includegraphics{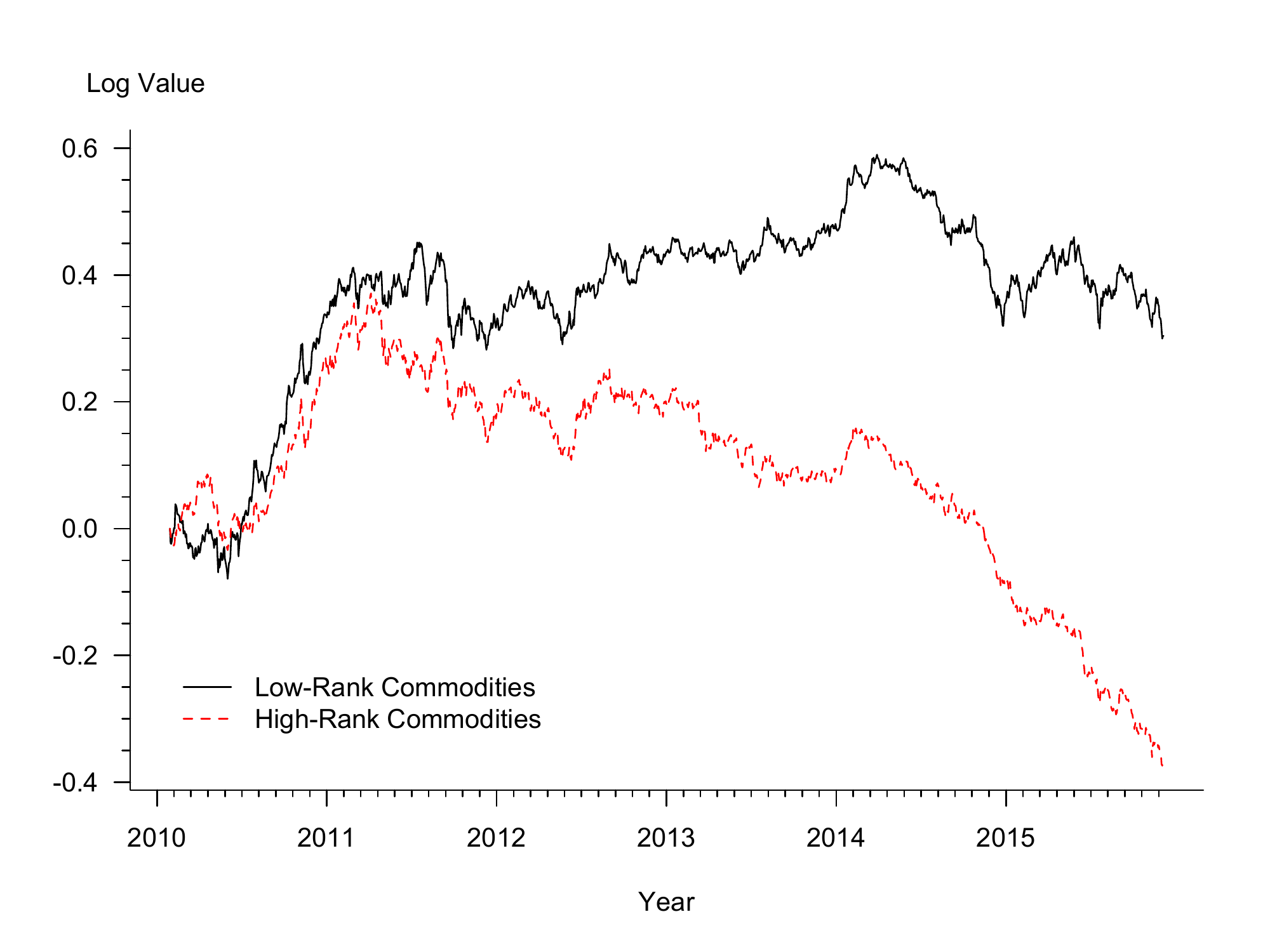}}}
\end{center}
\vspace{-24pt} \caption{Log returns for lower-ranked (bottom half) commodity futures and higher-ranked (top half) commodity futures portfolios, 2010 -- 2015.}
\label{sizeFigFutures3}
\end{figure}

Figures \ref{sizeFigRecent1} and \ref{sizeFigRecent2} show large and significant excess returns of lower-ranked commodities over higher-ranked commodities, much like the excess returns for lower-ranked commodity futures shown in Figures \ref{sizeFigFutures1} -- \ref{sizeFigFutures4} and Table \ref{regressionsTab}. Indeed, from 1980 -- 2015, the average annual excess return of the equal-weighted portfolio of lower-ranked commodities over the equal-weighted portfolio of higher-ranked commodities was 6.4\%. Furthermore, as \citet{Fernholz:2016d} shows, the correlation between these excess returns and monthly Russell 3000 stock returns is -0.13. Note that the rank effect excess returns shown in Figure \ref{sizeFigRecent1} are not driven by price declines for the highest-ranked commodity futures, which is in contrast to the excess returns shown in Figure \ref{sizeFigFutures3}. 
 
%Median sort monthly excess return of 0.47\% with a standard error of 0.15\%. Average annual excess return of 6.37\%.

\begin{figure}[ht]
\begin{center}
\vspace{-30pt}
\hspace{-15pt}\scalebox{0.66}{ {\includegraphics{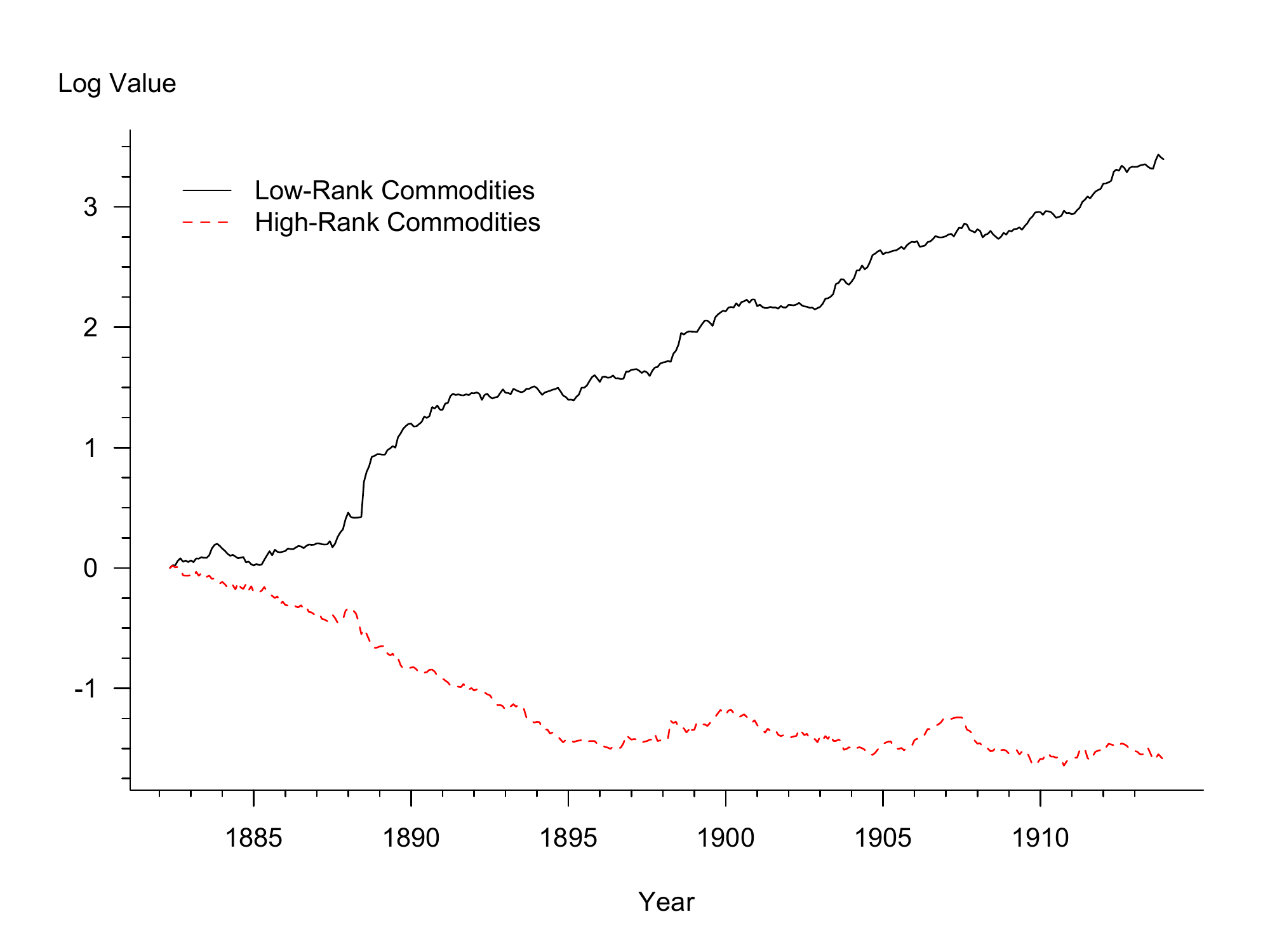}}}
\end{center}
\vspace{-24pt} \caption{Log returns for lower-ranked (bottom half) commodities and higher-ranked (top half) commodities portfolios, 1882 -- 1913.}
\label{sizeFigHistoric1}
\end{figure}

\begin{figure}[ht]
\begin{center}
\vspace{-5pt}
\hspace{-15pt}\scalebox{0.66}{ {\includegraphics{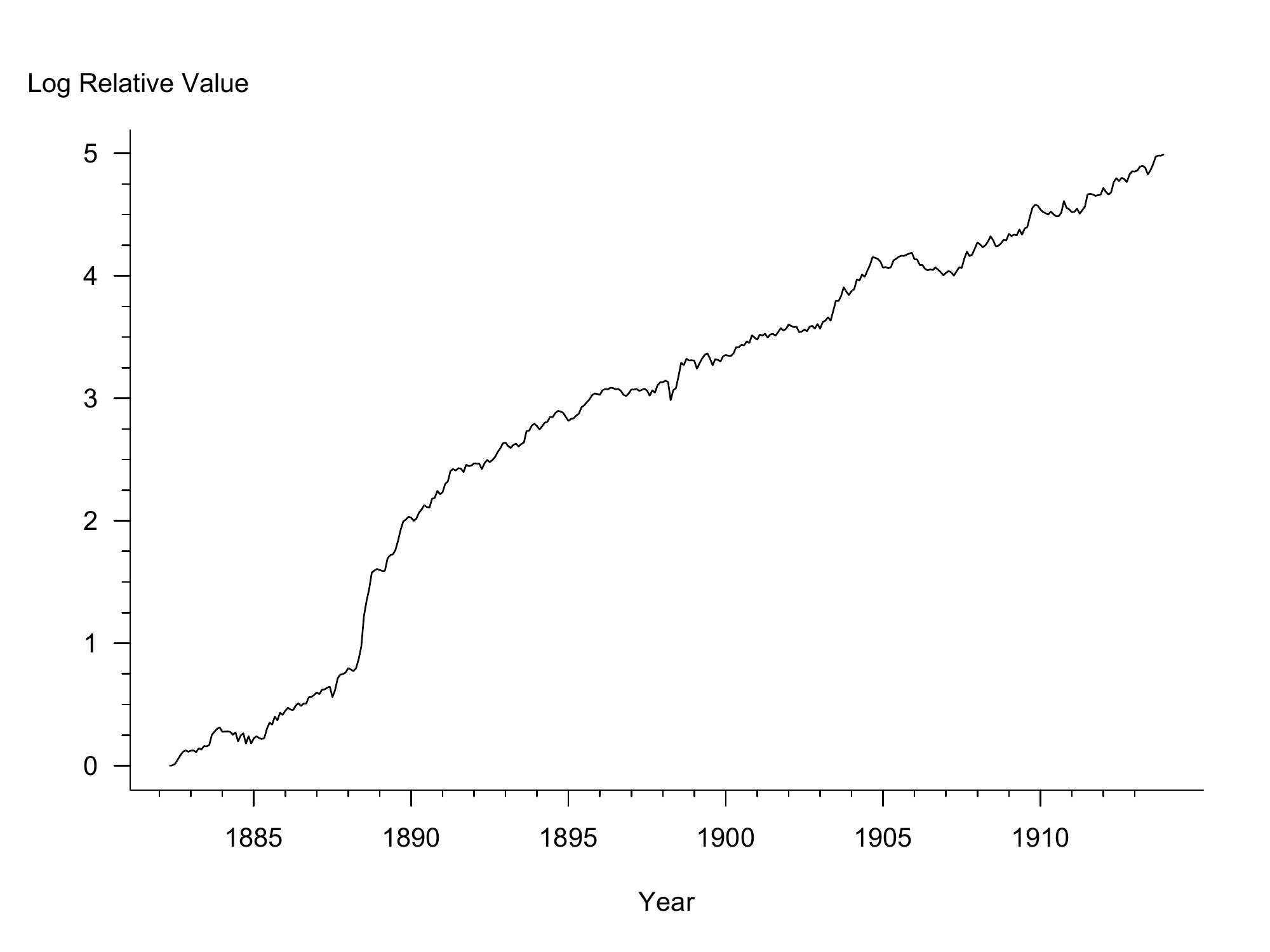}}}
\end{center}
\vspace{-24pt} \caption{Log return of lower-ranked (bottom half) commodities portfolio relative to higher-ranked (top half) commodities portfolio, 1882 -- 1913.}
\label{sizeFigHistoric2}
\end{figure}

Figure \ref{sizeFigHistoric1} plots the log value of an equal-weighted portfolio that invests in the bottom seven commodities together with the log value of an equal-weighted portfolio that invests in the top eight commodities for 1882 -- 1913. Figure \ref{sizeFigHistoric2} plots the log value of the bottom-seven commodities portfolio relative to the top-eight commodities portfolio from Figure \ref{sizeFigHistoric1}.\footnote{As with the portfolios from Figures \ref{sizeFigRecent1} and \ref{sizeFigRecent2}, we wait five months after the start date of January 1882 to begin trading so that the distribution of relative commodity prices has time to approach a stationary distribution and the rankings of commodities are meaningful.} Like with the spot commodity prices from 1980 -- 2015, these figures show large and significant excess returns of lower-ranked commodities over higher-ranked commodities. From 1882 -- 1913, the average annual excess return of the equal-weighted portfolio of lower-ranked commodities over the equal-weighted portfolio of higher-ranked commodities was 16.5\%.

%Median sort monthly excess return of 1.33\% with a standard error of 0.20\%. Average annual excess return of 16.47\%.

These results confirm the existence of further rank effects like the one we presented for commodity futures in Section \ref{futures}. In addition, these results highlight the robustness of our results across different commodity price normalization start dates. As discussed earlier, it is necessary to normalize commodity prices by equalizing them in the initial period in order to be able to compare and rank these prices in an economically meaningful way. Commodities are sold in different units, and hence their unnormalized spot and futures prices cannot be meaningfully compared. Notably, the normalization start date is irrelevant for the existence of a rank effect for commodities, as implied by the stationarity properties of relative commodity prices discussed in Section \ref{econom}. It is reassuring yet unsurprising to see this lack of reversal using three different data sets that span two centuries.

\subsection{Implications}
\begin{figure}[ht]
\begin{center}
\vspace{-5pt}
\hspace{-15pt}\scalebox{0.66}{ {\includegraphics{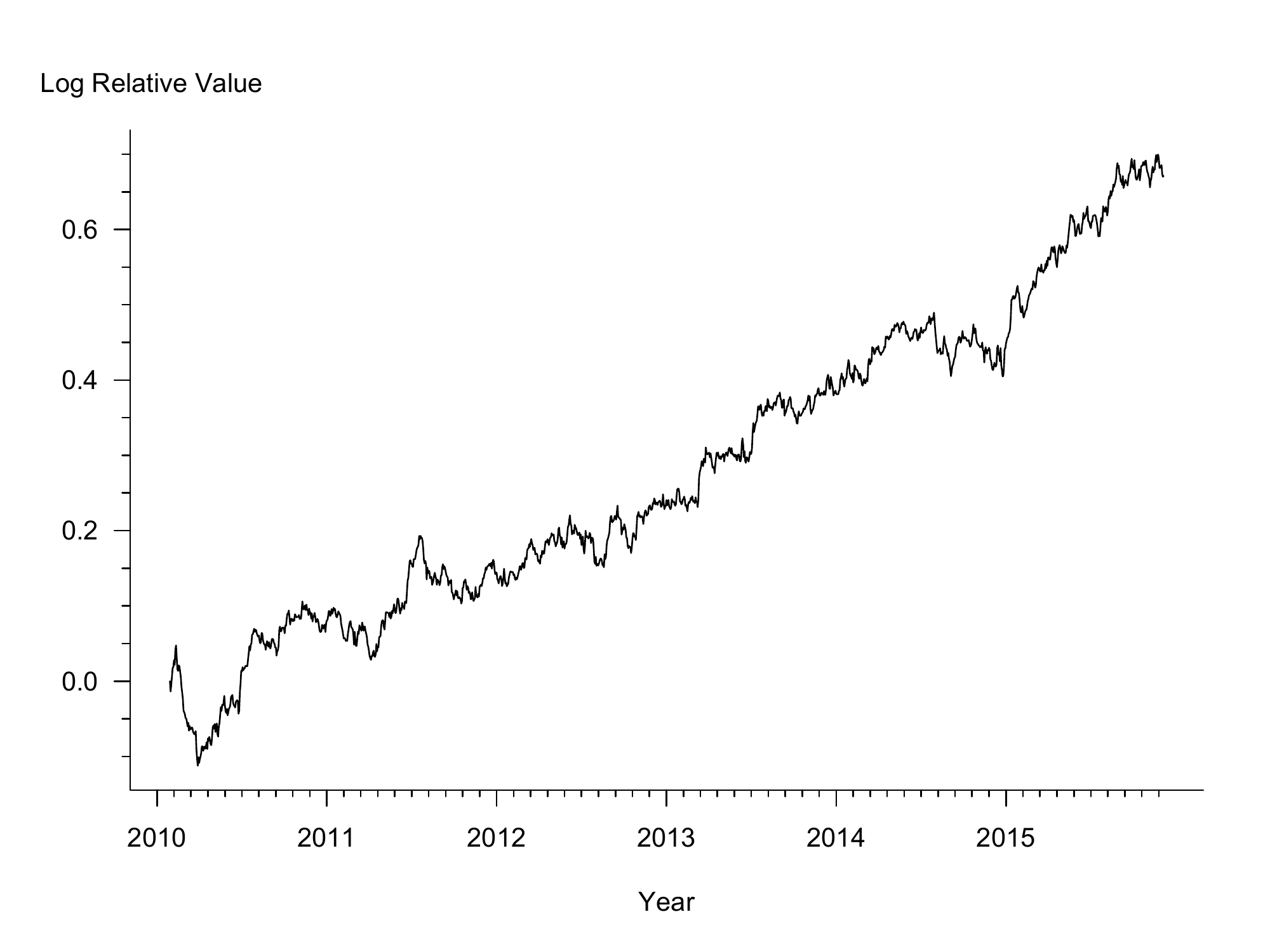}}}
\end{center}
\vspace{-24pt} \caption{Log returns for lower-ranked (bottom half) commodity futures relative to higher-ranked (top half) commodity futures portfolio, 2010 -- 2015.}
\label{sizeFigFutures4}
\end{figure}

Figures \ref{sizeFigFutures1} -- \ref{sizeFigFutures4} and Table \ref{regressionsTab} present large, tradable, and economically significant rank effect abnormal excess returns for commodity futures from 2010 -- 2015. These figures and tables also show that these LMH excess returns, in which lower-priced, lower-ranked commodities outperform higher-priced, higher-ranked commodities, are not explained by market risk. 

The sheer size of this tradable rank effect---average excess returns of more than 23\% per year and a Sharpe ratio nearly twice as high as the U.S. stock market---is notable. The fact that these returns are generated by simple, easy-to-construct portfolios that place equal weight on each commodity futures contract is all the more striking. Indeed, it is reasonable to expect that a more sophisticated trading strategy that optimizes the portfolio weights for the lower-ranked and higher-ranked commodities will do even better with even less market correlation.

In Section \ref{econom}, we demonstrate that there is nothing anomalous about the rank effect asset pricing anomaly. In fact, the excess returns shown in Figures \ref{sizeFigFutures1} -- \ref{sizeFigFutures4} and Table \ref{regressionsTab} are predicted by our nonparametric econometric theory. Furthermore, this theory suggests that there is no obvious or simple way for the LMH rank effect to be arbitraged away. After all, regardless of the rationality or irrationality of investors, commodity prices must always fluctuate and differ in a way that allows them to be ranked. We return to these fundamental and challenging questions in the next section.

%Figures \ref{sizeFigRecent1} -- \ref{sizeFigHistoric2} confirm the existence of a similar rank effect abnormal excess return for commodity spot prices from 1980 -- 2015 and 1882 -- 1913. We also show that this LMH excess return, in which lower-priced, lower-ranked commodities outperform higher-priced, higher-ranked commodities, is not explained by market returns. 

%\vfill
\vskip 50pt
%\pagebreak

\section{A Structural Explanation} \label{econom}

The large and economically significant excess returns described in the previous section are not a coincidence. In fact, these excess returns are predicted by and firmly grounded in economic and econometric theory. Using  nonparametric econometric methods, we show in this section that a rank effect in which lower-ranked, lower-priced commodities outperform higher-ranked, higher-priced commodities must exist in an economically and empirically realistic setting.

\subsection{Setup}
Consider an economy that consists of $N > 1$ commodities.\footnote{We follow the approach of and refer directly to \citet{Fernholz:2016c} for technical details and proofs.} Time is continuous and denoted by $t \in \rplu$, and uncertainty in this economy is represented by a filtered probability space $(\O, \F, \F_t, P)$. Let $\mathbf{B}(t) = (B_1(t), \ldots, B_M(t))$, $t \in [0, \infty)$, be an $M$-dimensional Brownian motion defined on the probability space, with $M \geq N$. We assume that all stochastic processes are adapted to $\{\F_t; t \in [0, \infty)\}$, the augmented filtration generated by $\mathbf{B}$.

The price of each commodity $i = 1, \ldots, N$ in this economy is given by the process $p_i$. Each of these commodity price processes evolves according to the stochastic differential equation
\begin{equation} \label{priceDynamics}
 d\log p_i(t) = \m_i(t)\,dt + \sum_{z=1}^M\d_{iz}(t)\,dB_z(t),
\end{equation}
where $\m_i$ and $\d_{iz}$, $z = 1, \ldots, M$, are measurable and adapted processes. The expected growth rates and volatilities, $\m_i$ and $\d_{iz}$, are general and practically unrestricted, having only to satisfy a few basic regularity conditions.\footnote{These conditions ensure basic integrability of equation \eqref{priceDynamics} and require that no two commodities' prices are perfectly correlated over time. See Appendix A of \citet{Fernholz:2016c}.} In particular, both growth rates and volatilities can vary across time and individual commodities in almost any manner.

Equation \eqref{priceDynamics} together with these regularity conditions implies that the commodity price processes in this economy are continuous semimartingales, which represent a broad class of stochastic processes \citep{Karatzas/Shreve:1991}. Furthermore, this analysis based on continuous semimartingales can also be extended to stochastic processes with occasional discrete jumps \citep{Shkolnikov:2011,Fernholz:2016a}.

The martingale representation theorem \citep{Nielsen:1999} implies that any plausible continuous process for commodity prices can be written in the nonparametric form of equation \eqref{priceDynamics}. Thus, our general setup is consistent with the equilibrium price dynamics that obtain in essentially any economic environment. Rather than committing to a specific model of commodity prices, we present general econometric results that are consistent with all models that satisfy some basic regularity conditions. Furthermore, our nonparametric approach nests more restrictive parametric statistical models of commodity price dynamics as special cases.

%If we think of firm size in terms of total market capitalization, then the firm size dynamics of equation \eqref{priceDynamics} correspond to stock price dynamics and hence capital gains. In this sense, our results in this section are valid only for stock market capital gains and not dividends. We shall return to the issue of dividends later in this section after presenting our main results about a size effect.

In order to examine the implications of different commodity price dynamics over time, it is useful to introduce notation for ranked commodity prices and commodity prices relative to the average price of all commodities in the economy. Let $p(t) = p_1(t) + \cdots + p_N(t)$ denote the total price of all commodities in the economy, and for $i = 1, \ldots, N$, let
\begin{equation} \label{shares}
 \theta_i(t) = \frac{Np_i(t)}{p(t)} = \frac{p_i(t)}{\frac{p(t)}{N}},
\end{equation}
denote the price of commodity $i$ relative to the average price of all commodities in the economy at time $t$. For $k = 1, \ldots, N$, let $p_{(k)}(t)$ represent the price of the $k$-th most expensive commodity at time $t$, so that
\begin{equation}
 \max (p_1(t), \ldots, p_N(t)) = p_{(1)}(t) \geq p_{(2)}(t) \geq \cdots \geq p_{(N)}(t) = \min (p_1(t), \ldots, p_N(t)),
\end{equation}
and let $\theta_{(k)}(t)$ be the relative price of the $k$-th most expensive commodity at time $t$, so that
\begin{equation} \label{rankShares}
\theta_{(k)}(t) = \frac{Np_{(k)}(t)}{p(t)}.
\end{equation}

In order to describe the dynamics of the ranked relative commodity price processes $\theta_{(k)}$, it is necessary to introduce the notion of a local time.\footnote{Local times are necessary because the rank function is not differentiable and hence we cannot simply apply \ito's Lemma.} For any continuous process $x$, the \emph{local time} at $0$ for $x$ is the process $\L_x$ that measures the amount of time the process $x$ spends near zero. We refer the reader to \citet{Karatzas/Shreve:1991} and \citet{Fernholz:2016c} for a formal definition of local times and a discussion of their connection to rank processes.

To be able to link commodity rank to commodity index, let $\o_t$ be the permutation of $\{1, \ldots, N\}$ such that for $1 \leq i, k \leq N$,
\begin{equation} \label{pTK}
 \o_t(k) = i \quad \text{if} \quad p_{(k)}(t) = p_i(t).
\end{equation}
This definition implies that $\o_t(k) = i$ whenever commodity $i$ is the $k$-th most expensive commodity in the economy. It is not difficult to show that for all $k = 1, \ldots, N$, the dynamics of the ranked relative commodity price processes $\theta_{(k)}$ are given by
\begin{equation} \label{rankSizeShareDynamics1}
  d\log\theta_{(k)}(t) = d\log\theta_{\o_t(k)}(t) + \frac{1}{2}d\L_{\log\theta_{(k)} - \log\theta_{(k + 1)}}(t) - \frac{1}{2}d\L_{\log\theta_{(k - 1)} - \log\theta_{(k)}}(t),
\end{equation}
a.s., with the convention that $\L_{\log \theta_{(0)} - \log \theta_{(1)}}(t) = \L_{\log \theta_{(N)} - \log \theta_{(N+1)}}(t) = 0$.\footnote{Throughout this paper, we shall write $dx_{\o_t(k)}(t)$ to refer to the process $\sum_{i = 1}^N1_{\{i = \o_t(k)\}}dx_i(t)$.} Together with equation \eqref{priceDynamics}, equation \eqref{rankSizeShareDynamics1} implies that\footnote{A formal derivation of equations \eqref{rankSizeShareDynamics1} and \eqref{rankSizeShareDynamics2} is provided by \citet{Fernholz:2016c}.}
\begin{equation} \label{rankSizeShareDynamics2}
\begin{aligned}
d\left(\log\theta_{(k)}(t) - \log\theta_{(k+1)}(t)\right) & =  \left(\m_{\o_t(k)}(t) - \m_{\o_t(k+1)}(t)\right)\,dt + d\L_{\log\theta_{(k)} - \log\theta_{(k + 1)}}(t) \\
& \qquad - \frac{1}{2}d\L_{\log\theta_{(k-1)} - \log\theta_{(k)}}(t) - \frac{1}{2}d\L_{\log\theta_{(k+1)} - \log\theta_{(k + 2)}}(t) \\
& \qquad + \sum_{z=1}^M\left(\d_{\o_t(k)z}(t) - \d_{\o_t(k+1)z}(t)\right)\,dB_z(t).
\end{aligned}
\end{equation}

\subsection{The Distribution of Relative Commodity Prices}
Let $\a_k$ equal the time-averaged limit of the expected growth rate of the price of the $k$-th most expensive commodity relative to the expected growth rate of all commodity prices together, so that
\begin{equation} \label{alphaK}
 \a_k = \limT1\intT\left(\m_{\o_t(k)}(t) - \m(t)\right)\,dt,
\end{equation}
for $k = 1, \ldots, N$.\footnote{Note that the growth rates in equation \eqref{alphaK} can vary over time and across any commodity-specific characteristics.} The relative growth rates $\a_k$ are a rough measure of cross-sectional mean reversion in commodity prices, since larger values of $-\a_k$ for the most expensive commodities imply faster mean reversion. In a similar manner, for all $k = 1, \ldots, N - 1$, let $\s_k$ be given by
\begin{equation} \label{sigmaK}
 \s^2_k = \limT1\intT\sum_{z=1}^M\left(\d_{\o_t(k)z}(t) - \d_{\o_t(k+1)z}(t)\right)^2\,dt.
\end{equation}
Note that $\s_k$ measures the volatility of the process $\log\theta_{(k)} - \log\theta_{(k + 1)}$, the relative prices of adjacent ranked commodities. Finally, for all $k = 1, \ldots, N$, let
\begin{equation} \label{kappaK}
 \k_k = \limT1\L_{\log\theta_{(k)} - \log\theta_{(k + 1)}}(T),
\end{equation}
and let $\k_0 = 0$ as well.

The \emph{stable version} of the process $\log\theta_{(k)} - \log\theta_{(k + 1)}$ is the process $\log\hat{\theta}_{(k)} - \log\hat{\theta}_{(k + 1)}$ defined by
\begin{equation} \label{stableVersion}
  d\left(\log\hat{\theta}_{(k)}(t) - \log\hat{\theta}_{(k+1)}(t)\right) = -\k_k\,dt + d\L_{\log\hat{\theta}_{(k)} - \log\hat{\theta}_{(k + 1)}}(t) + \s_k\,dB(t),
\end{equation}
for all $k = 1, \ldots, N - 1$.\footnote{For each $k = 1, \ldots, N$, equation \eqref{stableVersion} implicitly defines another Brownian motion $B(t)$, $t \in [0, \infty)$. These Brownian motions can covary in any way across different $k$.} The stable version of $\log\theta_{(k)} - \log\theta_{(k+1)}$ replaces all of the processes from the right-hand side of equation \eqref{rankSizeShareDynamics2} with their time-averaged limits, with the exception of the local time process $\L_{\log\theta_{(k)} - \log\theta_{(k + 1)}}$. By considering the stable version of these relative commodity price processes, we are able to obtain a simple characterization of the distribution of relative commodity prices.

The stable versions of the relative commodity price processes $\log\theta_{(k)} - \log\theta_{(k + 1)}$ only exist if the limits in equations \eqref{alphaK}-\eqref{kappaK} also exist. Note that the existence of these limits is a weaker condition than the existence of a stationary or steady-state commodity price distribution.

\begin{thm} \label{priceDistThm}
There is a stationary relative commodity price distribution in this economy if and only if $\a_1 + \cdots + \a_k < 0$, for $k = 1, \ldots, N - 1$. Furthermore, if there is a stationary relative commodity price distribution, then for $k = 1, \ldots, N - 1$, this distribution satisfies
\begin{equation} \label{priceDistEq}
 E\left[\log\hat{\theta}_{(k)}(t) - \log\hat{\theta}_{(k + 1)}(t)\right] = \frac{\s^2_k}{-4(\a_1 + \cdots + \a_k)}, \as
\end{equation}
\end{thm}

This theorem provides an analytic rank-by-rank characterization of the entire relative commodity price distribution.\footnote{We refer the reader to \citet{Fernholz:2016c} for a proof of Theorem \ref{priceDistThm}.} According to this characterization, an increase in the rate of cross-sectional mean reversion as measured by $-\a_k$ decreases the dispersion of commodity prices, while an increase in volatility $\s_k$ increases the dispersion of commodity prices. Note that Theorem \ref{priceDistThm} is valid under very weak assumptions, and hence it provides a characterization of essentially any steady-state relative commodity price distribution.

\begin{figure}[ht]
\begin{center}
\vspace{-30pt}
\hspace{-15pt}\scalebox{0.69}{ {\includegraphics{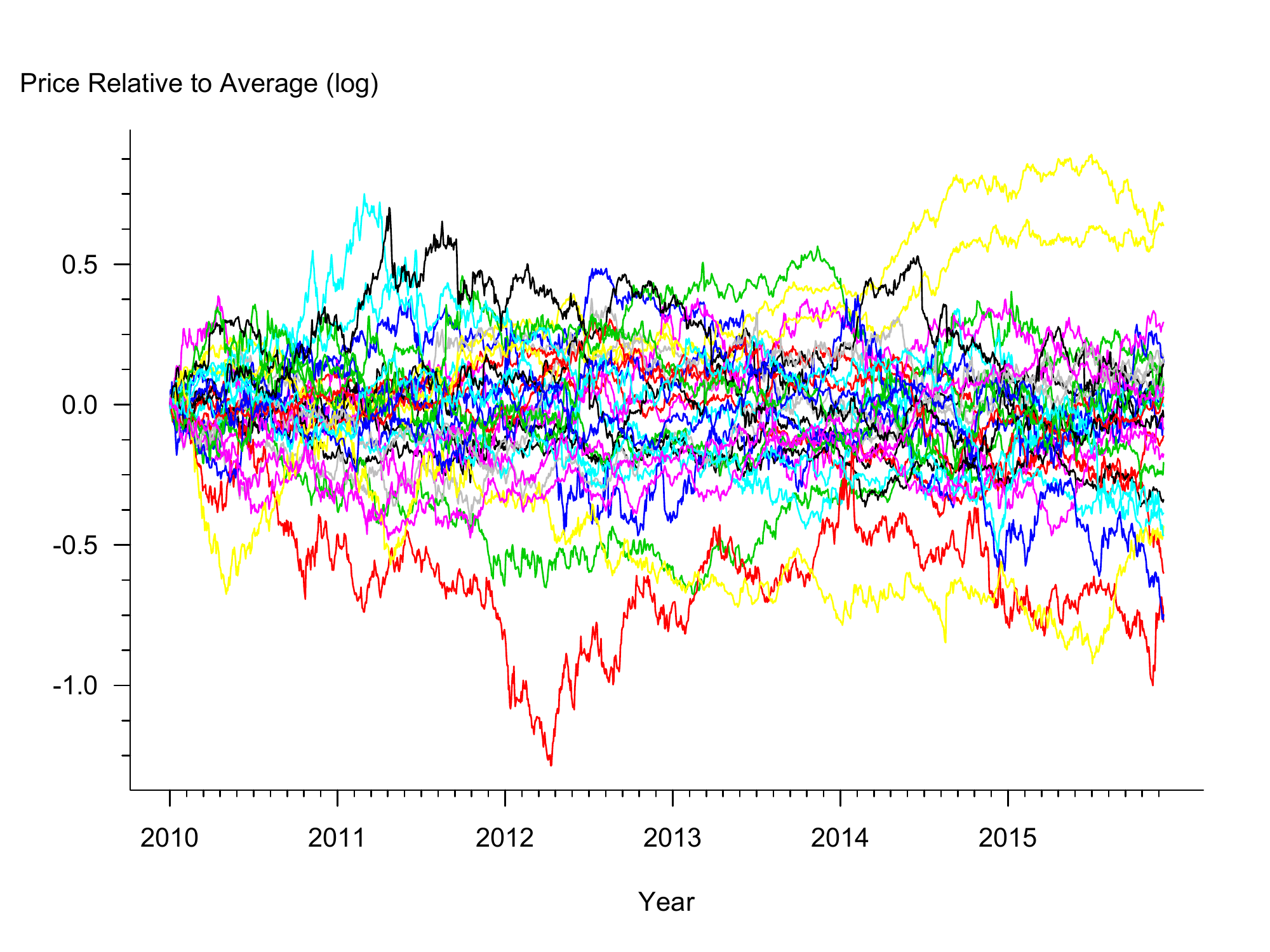}}}
\end{center}
\vspace{-24pt} \caption{Log commodity futures prices relative to the average, 2010 -- 2015.}
\label{pricesFigFutures1}
\end{figure}

\begin{figure}[ht]
\begin{center}
\vspace{-5pt}
\hspace{-15pt}\scalebox{0.69}{ {\includegraphics{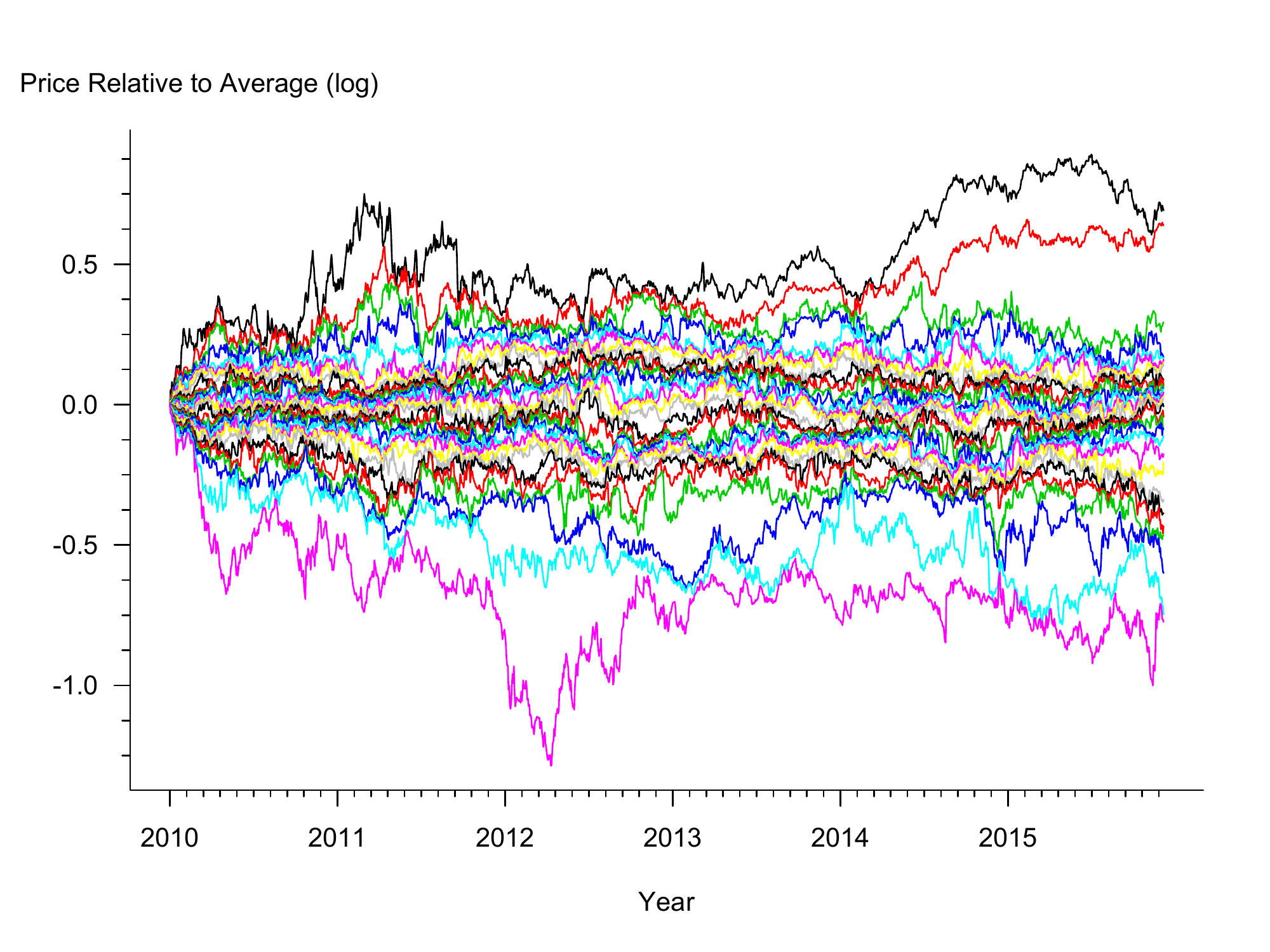}}}
\end{center}
\vspace{-24pt} \caption{Log ranked commodity futures prices relative to the average, 2010 -- 2015.}
\label{pricesFigFutures2}
\end{figure}

Figures \ref{pricesFigFutures1} and \ref{pricesFigFutures2} plot the log value of all commodity futures prices relative to the average from 2010 -- 2015 for both individual and ranked commodity futures. As the figures show, these relative commodity futures prices appear to be roughly stationary over time and hence we should expect that Theorem \ref{priceDistThm} will provide an accurate description of the observed relative commodity price distribution during this time period. Note that \citet{Fernholz:2016c} shows that this is true for spot commodity prices from 1980 -- 2015. As we explain in Section \ref{rankEffect}, the approximate stationarity of relative commodity prices over time implies that the rank effect excess returns we uncovered in Section \ref{sizeEffect} above are to be expected.

In addition to the empirical evidence, the basic economic notion of substitutability suggests that strong economic forces drive the long-run stability of relative commodity prices. It is difficult to imagine a plausible economic scenario in which any one commodity price grows arbitrarily larger than all other commodity prices. As a consequence, the stationarity of relative commodity prices evidenced by Figures \ref{pricesFigFutures1} and \ref{pricesFigFutures2} is not surprising. Instead, it is the far-reaching implications of this stationarity that are surprising.

\subsection{The Rank Effect} \label{rankEffect}
Theorem \ref{priceDistThm} states that there is a stationary distribution of relative commodity prices if and only if $\a_1 + \cdots + \a_k < 0$. In other words, the growth rates of the higher-ranked, higher-priced commodities in the economy must on average be lower than the growth rates of the lower-ranked, lower-priced commodities in the economy. This necessary condition is essentially a mean-reversion condition.

Our empirical results in Section \ref{sizeEffect} together with the mean-reversion condition of Theorem \ref{priceDistThm} lead to an important conclusion: there is nothing anomalous about the rank effect anomaly we uncover in this paper. The excess returns of lower-priced, lower-ranked commodities over higher-priced, higher-ranked commodities are exactly what is predicted for a stationary relative price distribution. In fact, even in the absence of stationarity, rank effect excess returns are still implied by long-run non-degeneracy of relative commodity prices, an extremely weak condition \citep{Fernholz:2016d}. 

The sheer size and simplicity of the rank effect excess returns presented in Section \ref{sizeEffect}---more than 23\% per year on average and a Sharpe ratio nearly twice as high as the U.S. stock market from 2010 -- 2015 using equal-weighted low-rank and high-rank portfolios---implies that there are enormous financial incentives for investors to exploit them. Regardless of the incentives, however, our structural econometric explanation for the rank effect makes it unclear how these excess returns can be arbitraged away by profit-seeking investors. This is arguably the most surprising feature of our results, and it is notably absent from other findings in this literature. Any plausible theory of asset price dynamics, whether rational or behavioral, implies that commodity prices will fluctuate and differ in a way that allows them to be ranked. As long as this is true, the rank effect for commodities is a necessary consequence of our econometric results. If we are willing to accept this implication, then the only way to reconcile our econometric and empirical results with the efficient markets hypothesis is to conclude that there must be a systematic relationship between rank and risk, a point emphasized and examined by \citet{Fernholz:2016d}. This relationship, however, is not supported by the market-neutrality of the rank effect excess returns reported in Table \ref{regressionsTab}.

The mean-reversion condition from Theorem \ref{priceDistThm} is related to the well-known size effect for stocks. Starting with \citet{Banz:1981}, a number of studies have shown the tendency for U.S. stocks with high total market capitalization---big stocks---to underperform U.S. stocks with low total market capitalization---small stocks. Although a number of potential explanations for this size effect have been proposed \citep{Fama/French:1993,Acharya/Pederson:2005}, our econometric theory offers a novel interpretation. According to Theorem \ref{priceDistThm}, the prices of higher-ranked assets must necessarily grow more slowly than the prices of lower-ranked assets, as long as the relative price distribution is stationary. If we apply this result to the size of stocks, then it states that higher-ranked, bigger stocks will yield lower capital gains than lower-ranked, smaller stocks on average over time. In other words, there will be a size effect. 

Of course, the size effect for stocks is complicated by the fact that stocks pay dividends and are subject to entry and exit, neither of which applies to commodities and hence is not taken into account in our empirical framework.\footnote{See \citet{Fernholz:2016e} for a detailed analysis of the size effect for stocks using an econometric framework similar to this paper.} Nonetheless, the fact that our simple econometric results both predict the rank effect for commodities we documented in Section \ref{sizeEffect} and offer an alternate, structural econometric explanation for the size effect for stocks is notable. In this sense, the size effect for stocks can be viewed as a special case of the general rank effect implied by the mean-reversion condition of Theorem \ref{priceDistThm}. 

The excess returns we uncover in this paper are also related to the well-known momentum factor proposed by \citet{carhart:1997}. Because the rank effect involves low ranks outperforming high ranks, these excess returns are in many ways the opposite of momentum excess returns. All else equal, it is natural to expect that an asset that performed well over the previous year will on average occupy a higher rank than an asset that performed poorly.\footnote{This prediction is less clear when considering the past performance of an asset over intermediate rather than recent horizons, as in \citet{novy-marx:2012}. An interesting direction for future research might be to link these intermediate-horizon momentum excess returns to the rank effect excess returns we uncover in this paper.} In any case, our findings unmistakably point to large and economically significant rank effect excess returns, regardless of how those returns relate to momentum.

%Should we say that reference Novy-Marx (2012??) again, and Asness, Moskowitz, and Pedersen (2013??)??

%including recent contributions such as the intermediate-horizon momentum anomaly of Novy-Marx (2012??) and the value and momentum anomalies of Asness, Moskowitz, and Pedersen (2013??). 

\vskip 50pt
%\vfill
%\pagebreak

\section{Conclusion} \label{conclusion}

Using spot and futures commodity price data spanning more than a century, this paper uncovers a large and economically significant rank effect in which lower-priced, lower-ranked commodities earn higher returns than higher-priced, higher-ranked commodities. We show that there is nothing anomalous about this anomaly. In fact, a nonparametric econometric framework predicts such a rank effect as a necessary consequence of a stationary relative asset price distribution. The sheer size of the tradable rank effect for commodity futures---more than 23\% per year on average and a Sharpe ratio nearly twice as high as the U.S. stock market from 2010 -- 2015---together with its low correlation with market returns is notable. After all, the foundational asset pricing theories of \citet{sharpe:1964} and \citet{Lucas:1978a} state that such large and predictable excess returns must be compensation for some kind of risk.

The rank effect for commodity prices raises fundamental and difficult questions for economics and finance. One interpretation of the structural econometric explanation we provide is as a foundation for a general asset pricing factor that can explain multiple anomalies. The well-known size effect for stocks, for example, is predicted by our econometric results. Are there multiple anomalies that are special cases of a more general rank effect that is a necessary consequence of a stationary distribution of relative asset prices? 

Perhaps the most compelling questions relate to the seeming robustness of the rank effect across any plausible investor behavior. Indeed, it is not clear how the rank effect for commodities can be arbitraged away since its existence relies only on the ability to compare and rank prices together with minimal and realistic assumptions for long-run relative commodity price dynamics. This striking conclusion suggests that the efficient markets hypothesis can be reconciled with our findings only if there is a systematic relationship between rank and risk, a point discussed in detail by \citet{Fernholz:2016d}. Finally, our results imply that traditional tests of market efficiency that impose unpredictability of returns using size or other rank-based predictors are often fundamentally flawed, since our econometric theory requires that rank predict returns in any closed and asymptotically non-degenerate asset market.

%Is the rank effect simply a general asset pricing factor, i.e. does our structural econometric explanation of the rank effect provide a foundation for many asset pricing anomalies like the well-known size effect for stocks? How is this arbitraged away, i.e. all asset prices (not just commodities) should be expected to randomly fluctuate in a way that allows them to be compared and ranked. But substitutability argues for a long run stability or stationarity, which implies a rank effect which contradicts unpredictability of excess returns. How do we reconcile this? Maybe traditional tests of market efficiency based on return predictability using size or other rank-like features are fundamentally flawed?

\vskip 50pt
%\vfill
%\pagebreak

%\begin{spacing}{1.1}

%\appendix
%\section{Proofs} \label{proofs}

%\begin{proofThm}

%\end{proofThm}

%\end{spacing}

%\vfill
%\vskip 70pt
\pagebreak

\begin{spacing}{1.2}

\bibliographystyle{chicago}
\bibliography{econ}

\end{spacing}

\pagebreak

\begin{table}[ht]
\vspace{15pt}
\begin{center}
\setlength{\extrarowheight}{3pt}
\begin{tabular} {|c||c|c|}

\hline

     & \phantom{Russell 3000 Daily Return ($\b$)} & \phantom{Russell 3000 Daily Return ($\b$)} \\ [-0.55cm]
Rank Cutoff     & Intercept         & Russell 3000 Daily Return ($\b$) \\ [0.06cm]

\hline

  15           &  4.24 \hspace{4pt} (1.73) &  0.075 \hspace{4pt} (0.016)   \\
   14          &   4.78 \hspace{4pt} (1.79) & 0.078 \hspace{4pt} (0.017)  \\
  13           &  5.75 \hspace{4pt} (1.87) &  0.087 \hspace{4pt} (0.018)  \\
  12           &  5.80 \hspace{4pt} (1.96) &  0.092 \hspace{4pt} (0.019)  \\
   11          &  6.59 \hspace{4pt} (2.05) & 0.093 \hspace{4pt} (0.019)  \\
   10         &  7.11 \hspace{4pt} (2.18) & 0.102 \hspace{4pt} (0.021)  \\
  9            &  7.13 \hspace{4pt} (2.29) & 0.109 \hspace{4pt} (0.022)  \\
  8            & 7.50 \hspace{4pt} (2.44) &  0.109 \hspace{4pt} (0.023)  \\
  7            &  9.35 \hspace{4pt} (2.67) &  0.107 \hspace{4pt} (0.025)  \\
   6           &  9.24 \hspace{4pt} (2.88) &  0.104 \hspace{4pt} (0.027)  \\
  5            &  9.32 \hspace{4pt} (3.19) &  0.092 \hspace{4pt} (0.030)  \\
  4            &  8.06 \hspace{4pt} (3.60) &  0.076 \hspace{4pt} (0.034)  \\
   3           &  8.08 \hspace{4pt} (4.11) &  0.040 \hspace{4pt} (0.039)  \\
   2           &  7.74 \hspace{4pt} (5.25) &  0.018 \hspace{4pt} (0.050)  \\
  1           &  10.90 \hspace{1pt} (7.82) &  -0.066 \hspace{2pt} (0.074)  \\

\hline

\end{tabular}
\end{center}
\vspace{-5pt} \caption{Regression results using daily excess returns (basis points) of lower-ranked commodity futures portfolios relative to higher-ranked commodity futures portfolios for all 15 rank cutoffs, 2010 -- 2015. For example, the third row from the bottom presents a typical rank-decile sort, which compares a portfolio consisting of the bottom three ranked commodity futures to the top three ranked commodity futures. Standard errors are reported in parentheses.} \label{regressionsTab}
\end{table}

\end{document}

%% file: mathdef.tex
\usepackage{amsmath}
\usepackage{amstext}
\usepackage{amsbsy}
\usepackage{amsthm}
\usepackage{amsfonts}
\usepackage{amssymb}
\usepackage{amscd}
\usepackage{eucal}

\theoremstyle{plain}
\newtheorem{thm}{Theorem}[section]

\theoremstyle{definition}

\def\ito{It{\^o}}

\def\footnoterule{\hrule \kern2.6pt}

\def\intT{\int_0^T}

\def\rplu{[\,0,\infty)}

\def\as{\quad\text{\rm a.s.}}
\def\a{\alpha}
\def\b{\beta}

\def\l{\lambda}
\def\L{\Lambda}
\def\r{\rho}
\def\s{\sigma}
\def\k{\kappa}
\def\d{\delta}
\def\t{\tau}

\def\o{\omega}
\def\O{\Omega}

\def\m{\mu}

\def\empty{\varnothing}
\def\0T{[\,0,T]}

\def\limT#1{\lim_{T\to\infty}\frac{#1}{T}}

\def\F{\CMcal{F}} 